\documentclass[
reprint,
 amsmath,amssymb,
 aps,
]{revtex4-2}

\usepackage{graphicx}% Include figure files
\usepackage{dcolumn}% Align table columns on decimal point
\usepackage{bm}% bold math

\begin{document}

\preprint{APS/123-QED}

\title{Topological Photonic Systems: Virtuous Platforms to Study \\ Topological Quantum Matter}

\author{Nitish Kumar Gupta}
\email{nitishkg@iitk.ac.in}
\author{Arun M. Jayannavar}

\affiliation{Centre for Lasers \& Photonics, Indian Institute of Technology Kanpur, 208016, India}
\affiliation{Institute of Physics, Bhubaneswar, Odisha 751005, India}

\date{\today}% It is always \today, today,

\begin{abstract}
 Topological insulators are a new class of materials that have engendered considerable research interest among the condensed matter community owing primarily to their application prospects in quantum computations and spintronics. Many of the associated phenomena, however, can be well reproduced in classical photonic systems with the additional advantage of relatively less demanding fabrication and engineered system characteristics. Therefore, the photonic analogs of topological materials have gained prominence in the last decade, not only in the field of optics but as an active research front of the topological physics at large. In this article, we succinctly review the fundamental concepts of topological physics and provide a concise description of the photonic topological insulators. 

\end{abstract}
\maketitle

\section{Introduction}

Although the studies in condensed matter systems encompass almost every aspect of macroscopic material behavior, over the years, we have come to realize that a large body of the noted phase transition phenomena in such systems can be approached systematically by relying on just two fundamental notions: Symmetry and Topology.  Thus, it is no surprise that when encountered with new observations, we look for their possible explanations in terms of these two paradigms. While a regimen underpinned with the concepts of symmetry and order parameter was established by Landau in his celebrated theory of symmetry-breaking~\cite{landau1936theory}, the investigations of topologically non-trivial behavior in condensed matter systems owe their genesis to the rather intriguing and surprising observations associated with quantum Hall effects and topological insulators~\cite{von2005developments,hasan2010colloquium}. Notwithstanding the fact that, in a relatively short span of time, we have been able to develop a working understanding of these relatively obscure aspects of the matter, attempts to uncover the full implications of topological order in quantum matter have a long road ahead. And it suitably explains the surge of research activity in the domain with major research themes like symmetry-protected topological phases, symmetry-enriched topological phases, and topological semimetals. 
Even as the explorations of topological concepts in fermionic systems are still underway, the scientists realized that the origins of these effects lie in the wave nature of the electrons, and hence these effects should manifest themselves in any system with dominant wave properties. This logical extension seemed promising and promulgated the search for topological characteristics in classical wave systems. The outcomes produced by almost a decade-long association have been promising; not only our understanding of topological classifications has been enriched, but some entirely new classes of applications have also come to the fore.  In this regard, photonic systems, in particular, have gained enormous attention. So much so that, today, topological photonics is one of the most active research domains in optics and photonics, promising not only coveted functionalities for photonic integrated circuits designs and for quantum information processing but also lying at the very frontier of research in topological physics at large. The excessive focus on photonic topological systems also stems from the fact that as far as practical applications are concerned, the photonic counterparts of topological matter have some inherent advantages in terms of fabrication, making them a favored testbed for verifying many of the uncanny theoretical proposals and to realize working prototypes in accordance with the existing device architectures.

\section{A primer on topological matter}

Before moving on to discuss various intricacies related to the photonic topological insulators, we will succinctly revisit some of the relevant topological concepts in low-dimensional systems. For this purpose, we will rely on the paradigmatic one-dimensional (1D) model, called the Su-Schrieffer-Heeger (SSH) model~\cite{su1979solitons}, and elaborate on a few aspects of it.

Topology in modern mathematics refers to a concept where some property of an object is preserved under continuous deformation. The idea has been employed traditionally for the classification of geometrical entities in mathematics. To this objective, an important task is to define a justified and calculable parameter associated with the system that is capable of characterizing the possible topological configurations. This parameter is called the topological invariant, which, in the case of geometrical objects, is defined in terms of the Gauss-Bonnet theorem.
\begin{figure}[htbp]
\centering
\fbox{\includegraphics[width=\linewidth, height=2cm]{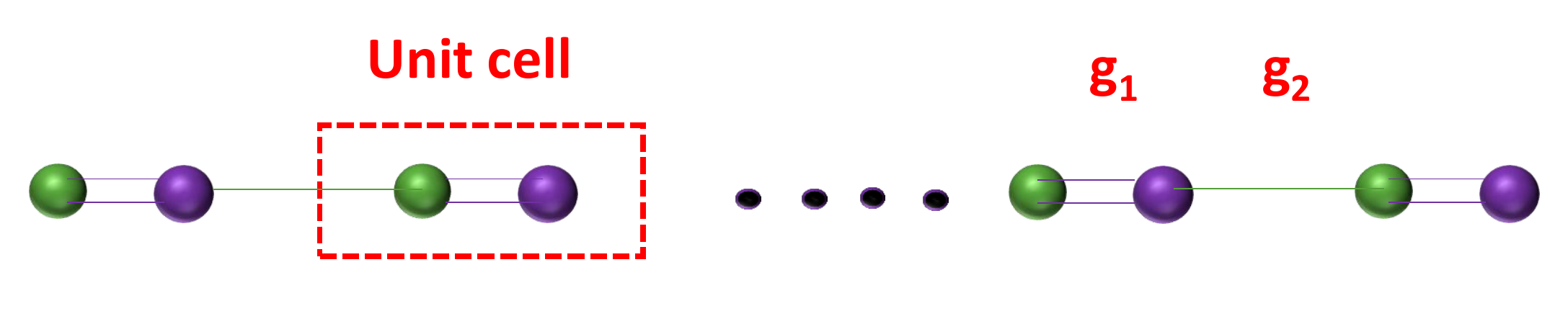}}
\caption{The setup of SSH model: $g_1$ represents the intra-unitcell hopping probability and $g_2$ represents the inter-unitcell hopping probability.}
\label{fig1}
\end{figure}
A few decades ago, on account of certain developments that were inexplicable in terms of Landau’s framework, it was recognized that in condensed matter systems also certain phases can be characterized by defining a suitable topological invariant. Topological invariants are quantized quantities (integers) that remain unchanged under arbitrary continuous deformations. Objects which possess the same topological invariant are called to be in the same topological phase (topologically equivalent objects). Any process that leads to a change in the topological invariant, is said to have initiated a topological phase transition, which apparently does not break any symmetry of the system. 

To better understand these aspects, we focus on 1D systems, which on account of the reduced complexity, allow us to keep our focus on some of the most prominent features. The model of choice in 1D is the celebrated SSH model, first proposed to describe the elementary excitations in conducting polymers. Along with its generalizations, it is among the first models to feature topological behavior. Nevertheless, the topological insulator described by this prototypical model captures the essential physics and hence forms a logical starting point of our study.  A short description of the SSH model is as follows: SSH model pertains to a tight-binding 1D chain with energy-dependent hopping probabilities. In the case of the periodic boundary conditions, the model results in two energy bands, the topological characteristics of which can be altered by changing the ratio of staggered hopping probabilities.

Figure 1 depicts an N-site SSH lattice with each unit cell consisting of two sublattice sites. The double bond represents the intra- unit cell hopping ($g_1$), and the single bond represents the inter-unit cell hopping ($g_2$). The Hamiltonian for the system can be written as: 

\begin{equation}
    \boldsymbol{H}= \sum_{j} ({g_1} {c^{\dagger}_{j,1}} {c_{j,2}}+ h.c.) -\sum_{j} ({g_2} {c^{\dagger}_{j,2}} {c_{j+1,1}}+ h.c.)
\end{equation}
\\

\begin{figure}[htbp]
\centering
\fbox{\includegraphics[width=\linewidth, height=6.3cm]{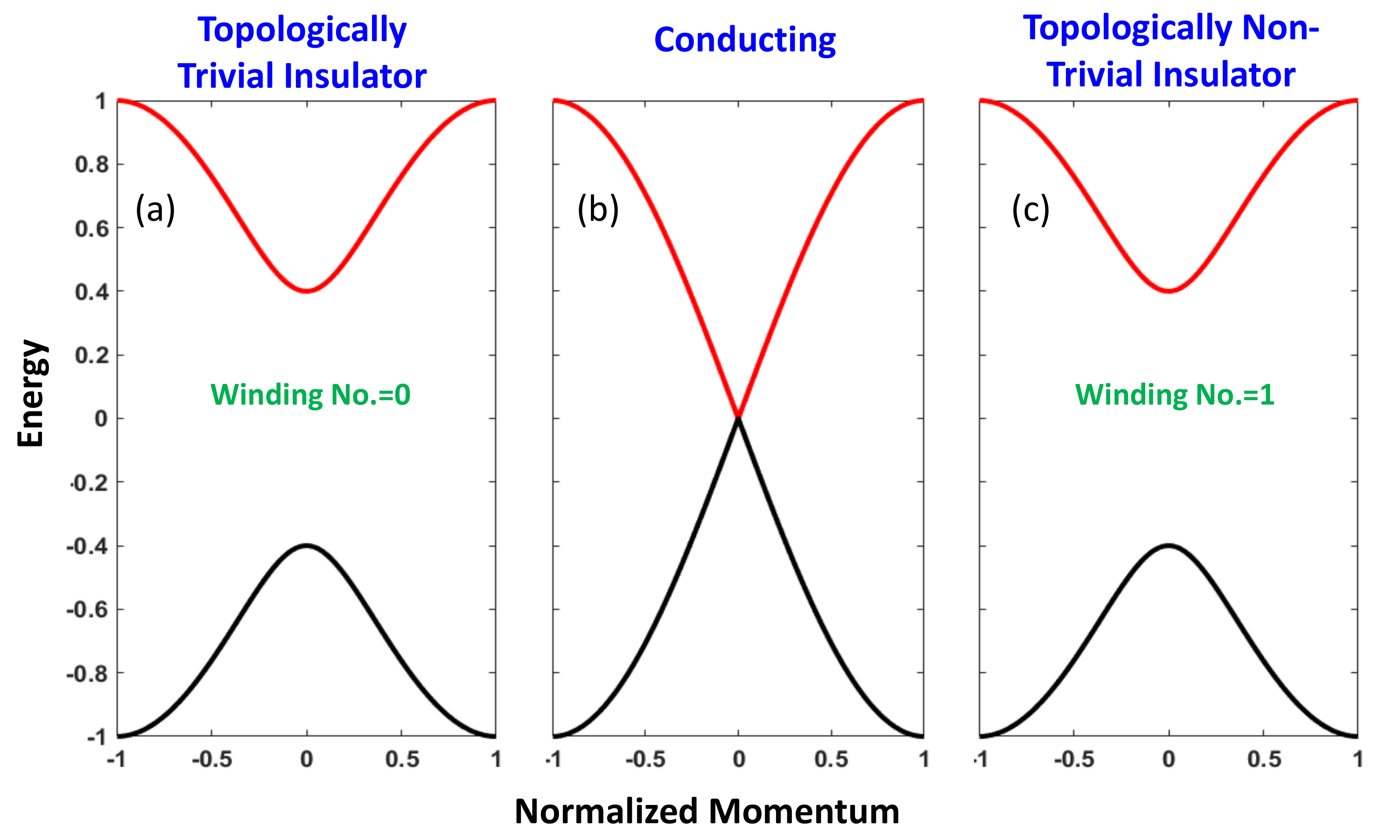}}
\caption{Calculated eigenvalue spectra for the SSH model: (a) $g_1>g_2$, (b) $g_1=g_2$, (c) $g_1<g_2$.}
\label{fig2}
\end{figure}

which gives $\boldsymbol{\mathcal{H}}(k)=$

\begin{equation}
\begin{aligned}
        \begin{pmatrix}
0 & g_1-g_2\cos(k)+ i g_2\sin(k)\\
g_1-g_2\cos(k)- i g_2\sin(k) & 0
\end{pmatrix}
\end{aligned}
\end{equation}

Hence, for a periodic lattice, the energy eigenvalues of the corresponding band Hamiltonian come out to be :

\begin{equation}
    E(k)= \pm\sqrt{g_1^2+g_2^2 \cos^2(k)-2g_1g_2\cos(k)+g_2^2\sin^2(k)}
\end{equation}

The corresponding band structures have been plotted in Figure 2. The topological invariant in this bulk band structure can be ascertained by calculating the winding number associated with the bands, the definition for which is given by:

\begin{equation}
    \boldsymbol{\mathcal{W}}= -\frac{1}{\pi}  \oint i\left\langle  \psi(k) \middle| \nabla_{k} \middle| \psi(k) \right\rangle dk 
\end{equation}

The winding numbers for the periodic SSH realization have been calculated and marked in green in the band diagrams of Figure-2, for three scenarios characterized by hopping probabilities. Although the bandstructures for Figure 2(a) and 2(c) look identical but the corresponding systems exist in distinct states, which becomes evident by calculating the corresponding winding numbers only.

Moving on to practical realizations, when we model a finite chain of atoms, the concept of bulk-edge correspondence necessitates the observation of topological edge states equal to the difference in their bulk winding numbers. This statement has been verified for a fifty-site SSH chain for which the complete system Hamiltonian is written, and numerically solved for the eigenvalue spectrum, as depicted in Figure 3.
In the spectrum, we observe the appearance of two zero-energy edge states, as predicted by the bulk-edge correspondence. The conditions of appearance of these edge states, as mentioned before, are determined by the relative strength of the two hopping amplitudes. In Figure 4, we also plot the eigenmode profiles corresponding to an edge mode and a bulk mode for a thirty-site SSH chain. These eignemode profiles sufficiently highlight the distinctive characteristics of the two sets of modes.

\begin{figure}[htbp]
\centering
\fbox{\includegraphics[width=\linewidth, height=6.2cm]{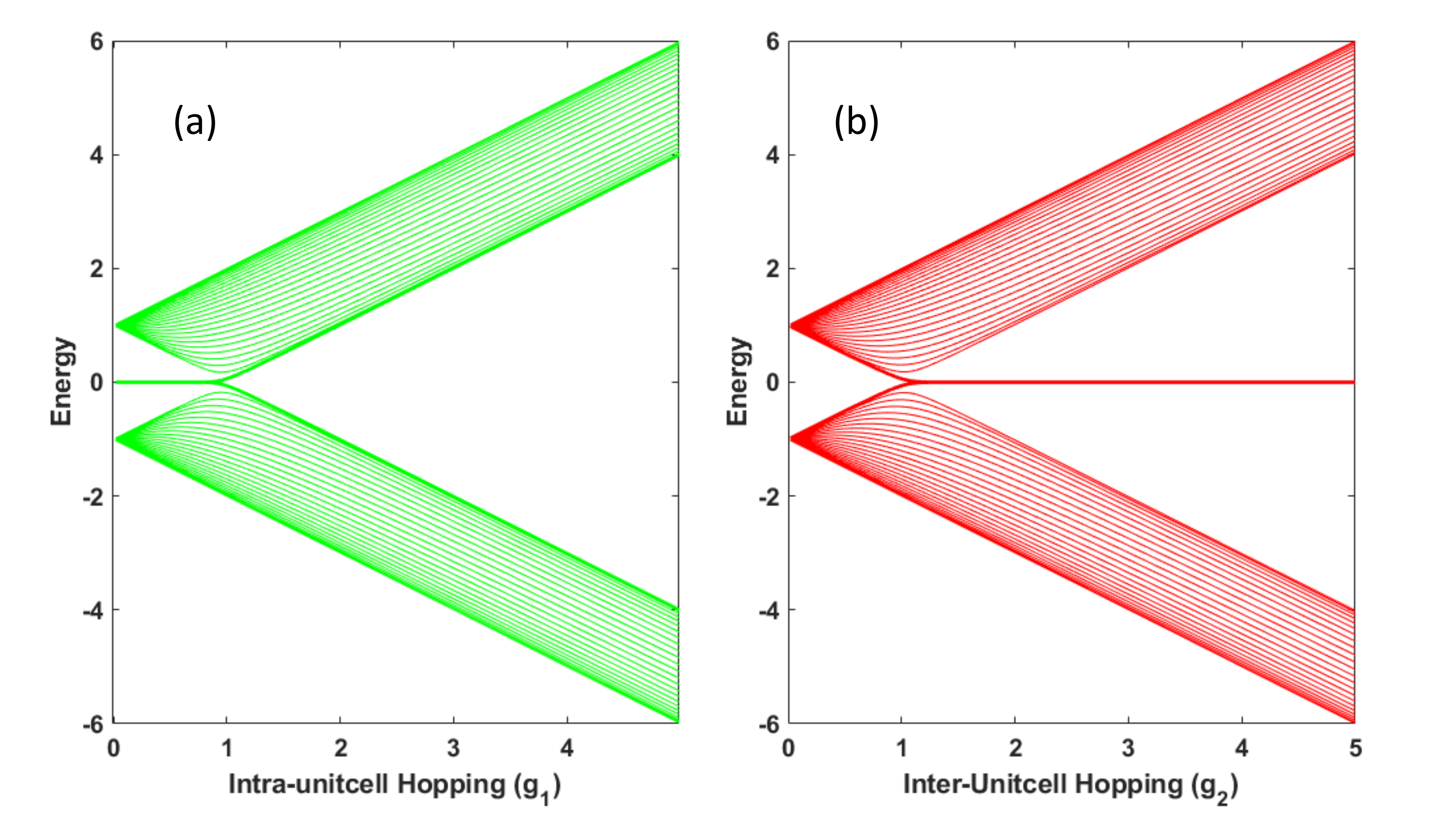}}
\caption{Calculated eigenvalue spectra for a finite SSH chain with N=50, from two perspectives : (a) $g_2$ is fixed while $g_1$ is varied, (b) $g_1$ is fixed while $g_2$ is varied.}
\label{fig3}
\end{figure}

Having been acquainted with few fundamentals of topological behavior, let us move on to the higher dimensional systems in a chronological sequence of their discoveries: one of the first systems where the topological aspects lead to inexplicable phenomena were the now-famous quantum Hall insulators. In these systems, the electrons confined to two dimensions (2D) form quantized cyclotron orbits (called Landau levels) under the application of a uniform magnetic field. When a measurement of conductance is carried out with the electron energy corresponding to the energy gap, the edge conductance turns out to be constant within an accuracy of around one part in a billion, regardless of the purity of the sample.

This was a breakthrough discovery that made us aware of the connections between conductance and topology and forced us to ponder upon the prospects that this new degree of freedom held for the future. Looking for an alternative system, not relying on Landau levels, in 1988, Haldane conceptualized a theoretical model consisting of periodic potentials with similar observations— the quantum anomalous Hall effect. Then came the proposal by Kane and Mele in 2005 about the spin Hall effect in Graphene~\cite{kane2005z}. The calculations in the paper suggested that under the influence of spin-orbit potential, the one-atom-thick layer of Graphite can behave like a quantum spin Hall insulator, doing away with the need for the external magnetic field. The idea was immediately carried forward to three dimensions (3D) in two separate works~\cite{fu2007topological,moore2007topological}. 
\begin{figure}[htbp]
\centering
\fbox{\includegraphics[width=\linewidth, height=5.5cm]{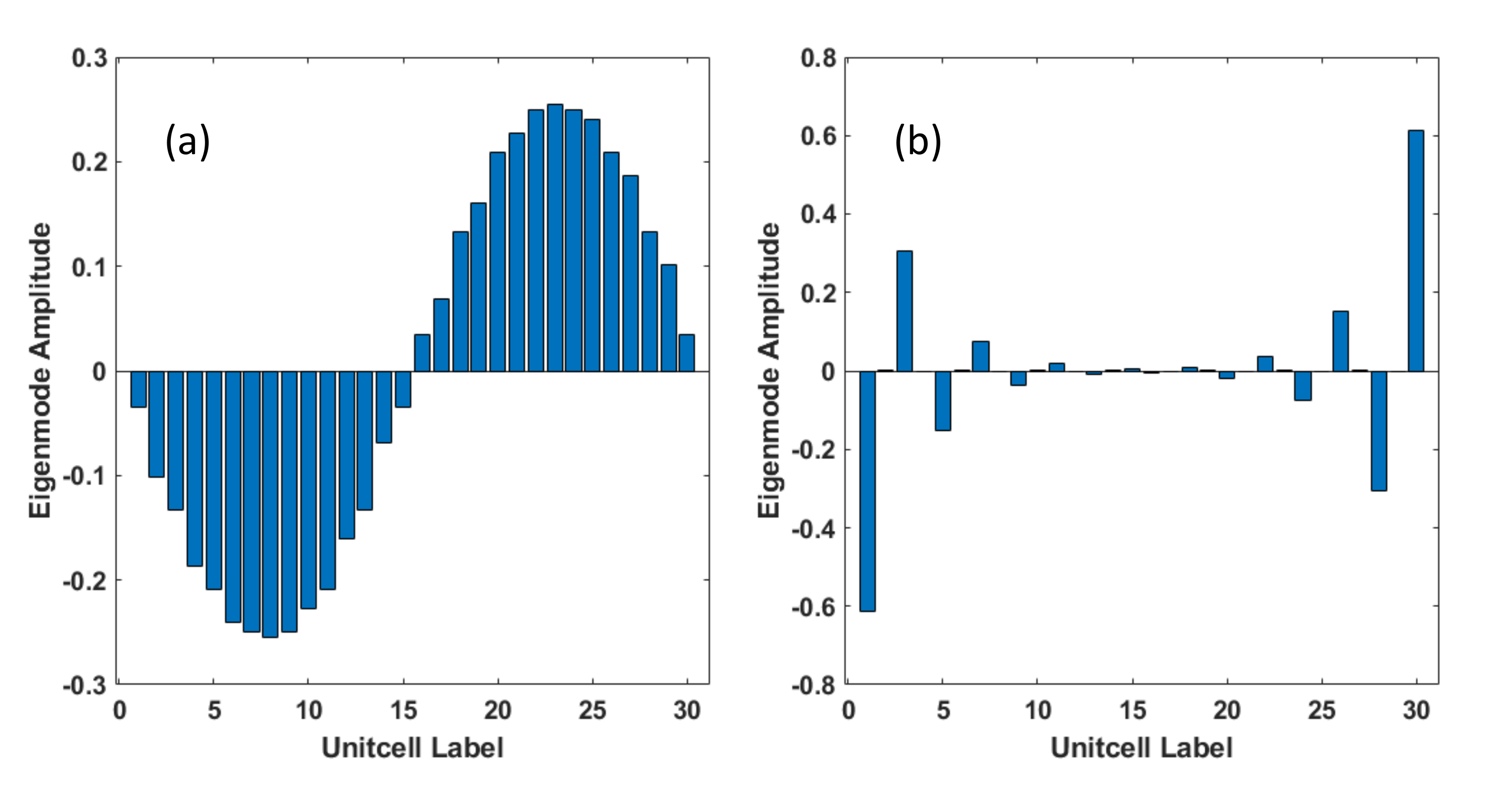}}
\caption{Calculated eigenmode profiles across the SSH chain for N=30, for two representative modes: (a) bulk mode, (b) edge mode.}
\label{fig4}
\end{figure}
The calculations in these works have shown that 3D blocks of matter can also exhibit such quantum effects. This paper by Moore and Balents~\cite{moore2007topological} also coined the term “topological insulator” for these peculiar states of matter.

\section{Photonic Analogues of Topological Insulators: }

The most remarkable characteristic of topological insulators is that these materials support gapless unidirectional spin-polarized states at their surface (in case of 3D) or at their edges (in 2D), leading to robust conduction properties. However, in most of the materials, the required spin-orbit coupling is too weak to engender the topological nature, and hence natural topological insulators are hard to find. That is why any prospects of artificial media with engineered topological characteristics become very exciting. Specifically, the photonic analogs of topological insulators can be designed with unprecedented control using artificial electromagnetic structures like metamaterials and photonic crystals. 
The first step in this direction was taken not long ago by conceptualizing a 2D electromagnetic bandgap material in which the appearance of topological properties was made possible by the Faraday effect. Indeed, the birth of topological photonics can be attributed to the 2008 article by Haldane and Raghu~\cite{haldane2008possible}, where they had suggested a mechanism to emulate the photonic analog of the quantum (anomalous) Hall effect. The proposal utilized the broken time-reversal symmetry in a nonreciprocal media to realize the signature one-way guided electromagnetic states associated with the quantum Hall realizations.  Soon after, the idea was realized by demonstrating the unidirectional waveguiding using the photonic chiral edge modes at microwave frequencies~\cite{wang2009observation}. Together it was an event that established a realistic possibility of creating light transport channels that offer a scattering free transport of electromagnetic waves even under the influence of fabrication imperfection and environmental changes. The idea bore the potential of significantly changing our perspective on light transport and scattering. Over the years, it has been tested and refined by associating it with almost all the existing photonic platforms, from metamaterials to the arrays of ring resonators, resulting in some exciting device applications discussed in the next section.

\section{The promise of topological photonics}

The immediate practical applications of topological photonics range from integrated photonic circuitry to slow light devices with arbitrarily large immunity to disorder. Furthermore, their peculiar characteristics also hold the key for industrial realization of quantum technologies and spintronic devices in the near future. Let us briefly look at some of the promising avenues whose genesis itself is attributable to the incorporation of topological aspects in photonic systems.

\subsection{Waveguides with backscattering immune propagation for integrated photonic circuitry}

Photonic chips promise a higher data processing speed and significantly lower losses than conventional electronics chips, potentially supporting high-capacity data routing requirements of the future. Although better than their electronics counterparts, the conventional photonic circuitry still suffers from the reduced efficiency of light transport and performance degradation due to the unwanted feedback caused by the backscattering events.  This issue is becoming a major concern in envisaging large-scale optical integration in photonic circuitry. Furthermore, as the requirements of miniaturization become more demanding, this problem will seriously limit the application potential. This is where the photonic topological insulators can help in a big way by offering the ideal transport properties. Beginning from the first demonstration of photonic topological waveguiding in microwave frequencies, topological photonics has been steadily moving towards delivering robust waveguiding in densely packed photonic integrated circuits by providing immunity to ambient perturbations and fabrication imperfections.
The intense effort for realizing fabrication conducive topological photonic waveguides can be categorized in three primary research directions- (1)	Topological photonic crystal waveguiding in photonic analogs of quantum Hall systems: these systems support chiral 1D edge states by utilizing external magnetic fields;
(2)	Topological photonic crystal waveguiding in photonic analog of quantum spin Hall systems: these systems support oppositely propagating helical edge states at the same boundary. The propagation direction of these helical edge states depends on their photonic pseudospin. Such systems have been realized using bianisotropic metamaterials; 
(3)	Topological photonic crystal waveguiding in the photonic quantum valley Hall system: these systems rely on breaking the spatial inversion symmetry to open a gap around the Dirac point of the system.

\subsection{Topological Photonics as a key Enabler for Quantum Technologies}
Quantum computers that rely on qubits for logic operations can theoretically outperform the existing computation platforms. However, the qubits, as envisaged initially, based on superconducting circuits and trapped ions are not immune to electromagnetic interference, apart from being highly demanding in terms of cryogenic temperatures of operation. On the other hand, due to the limited decoherence of photonic systems, Qubits relying on photons have many inherent advantages. 

Although more prolific than the superconducting qubits, the preservation of entanglement (which is necessary to make the quantum computers work) is a non-trivial task in photonic realization as well, as entanglement is very fragile. The celebrated topological protection is considered an apt solution to this problem as it can defend the photonic qubits from scattering and other disruptions (which may be caused due to external perturbations or fabrication errors). This can help significantly maintain the correlation and entanglement of photons over considerable distances, making sure that the quantum information does not get lost. This prospect has indeed proved to be a big impetus for research in topological photonics. 
Apart from minimizing the decoherence during photon propagation, topological photonics can also be instrumental in realizing stable quantum light sources.

\section{Future trends in topological photonics}

\subsection{Non-Hermitian Topological Photonics}

The conventional framework of quantum mechanics demands conservation of probabilities and unitary temporal evolution and hence is Hermitian by design in which every observable of the system is described by a Hermitian operator. Being Hermitian, the eigenvalues of the operators are real, and their eigenvectors are orthogonal.  A change in perspective, however, was brought about by the observations from Bender and Boettcher~\cite{bender1998real}. They noticed that a large class of non-Hermitian Hamiltonians can exhibit real eigenvalues provided that they satisfy the commutation relation with parity-time (PT) operator. To be precise, a PT-symmetric Hamiltonian exhibits a PT-symmetry Unbroken phase where it exhibits the real eigenvalue spectra. Beyond a point of degeneracy, it spontaneously enters the Broken PT-symmetry phase where the eigenvalues become complex, and the corresponding eigenvectors do not remain orthogonal or become skewed to each other.  The phase transition occurs when the non-hermiticity parameters of the system exceed a certain threshold and result in the appearance of non-Hermitian degeneracies. Thus, PT-symmetry offers conserved probabilities and physical observables without necessarily conforming to the Hermitian framework. Because of this, the PT-symmetric systems have been studied extensively in the last decade to derive new functionalities from the non-negligible losses in photonic systems. Apart from PT-symmetric systems, the non-Hermitian Hamiltonians, which do not commute with PT operator, are also taking center stage in photonic studies as the gain and loss are almost exclusively an integral part of the active photonic devices.

\begin{figure}[htbp]
\centering
\fbox{\includegraphics[width=\linewidth, height=5.5cm]{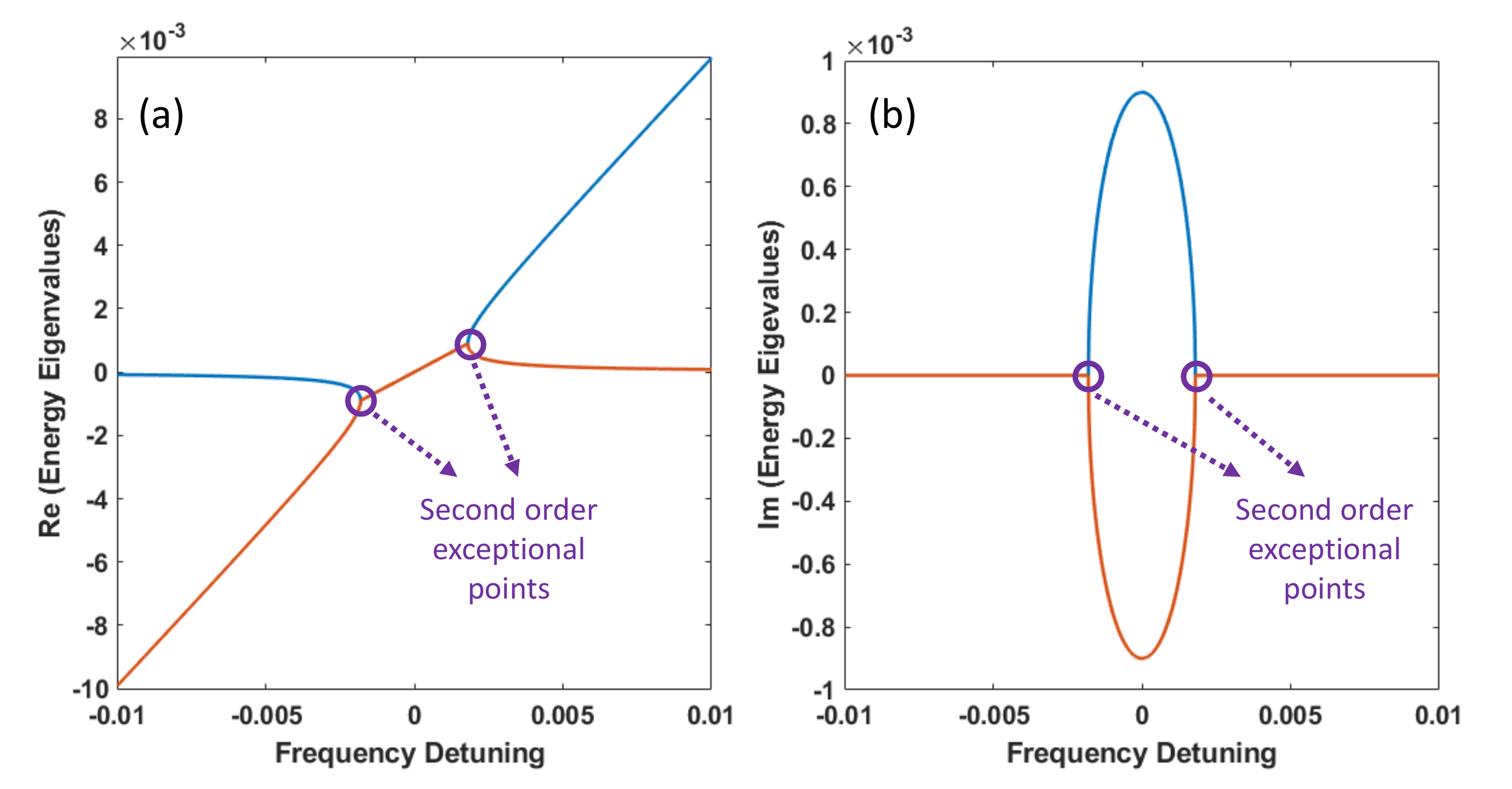}}
\caption{Occurrence of exceptional points with frequency detuning in a typical two-band non-Hermitian Hamiltonian.}
\label{fig5}
\end{figure}

As these new directions are emerging, enormous research activity has been noticed in the last five years regarding the topological aspects of the non-Hermitian systems. Although, first scientists grappled long enough with the question of whether it is reasonable to talk of topological properties in non-Hermitian systems, but now with the demonstrations of topological lasers, that question has taken a hind seat. The focus is now on combining these two phenomena to get some neoteric and exciting physics. Indeed, over a matter of few years, many surprises have been witnessed in the studies of non-Hermitian systems, such as the non-Hermitian skin effects and breakdown of bulk-edge correspondence. In the case of the non-Hermitian skin effect, the extended Bloch modes of the system also get localized at the edges or boundaries. This has created an immediate requirement of an all-encompassing framework that can explain the topological aspects peculiar to non-Hermitian systems, and its is a major reason behind extensive theoretical activity in this arena. Therefore, in the context of non-Hermitian systems, the first problem to address is to properly define topological invariants~\cite{leykam2017edge}.
Although the approach of generalizing the topological invariants corresponding to the Hermitian counterpart works satisfactorily in some instances (when the Hermiticity condition is substituted with the pseudo-Hermiticity or PT-symmetry condition), in some other instances, these topological invariants become inadequate. It has led to a vigorous search for neoteric frameworks and topological invariants in non-Hermitian systems that can reestablish the bulk-edge correspondence in these systems as well. Out of the many contributions, we will mention here two prominent proposals, which have contributed to our current understanding of this developing area of study: the first proposal tries to approach a generalized bulk-boundary correspondence by augmenting the Bloch Hamiltonian itself, which leads us to a notion of generalized Brillouin zone~\cite{yao2018edge,yao2018non}. This methodology has been termed as non-Bloch description as it defines non-Bloch topological invariants (Chern number), capable of predicting the existence of chiral edge states. In this manner, such a framework is more in tune with the familiar picture of conventional bulk-boundary correspondence (defined in terms of (conventional) topological invariants). The second approach, however, deviates considerably from the aforementioned setup, where a generalization has been approached in the form of biorthogonal bulk-boundary correspondence~\cite{kunst2018biorthogonal,edvardsson2019non}. This framework is based on biorthogonal quantum mechanics (which permits the treatment of non-Hermitian observables), where localization transitions of the biorthogonal wavefunctions are examined. Apart from these, various refinements and complementary approaches for the reestablishment of bulk-boundary correspondence have been put forth, which have been documented in the recent review~\cite{bergholtz2021exceptional}.

Furthermore, the topological aspects associated with the encirclement of an exceptional point of a non-Hermitian system have also been gaining attention. Let us understand the context of these studies: according to the adiabatic theorem, whenever a conservative (Hermitian) gapped Hamiltonian is evolved slowly and in a cyclic manner, the system returns to its initial state apart from a global phase pick up, termed as the Berry phase. The path along which this cyclic evolution is performed can have different topological properties depending on whether it encircles a singularity of the system or not. Particularly, in non-Hermitian systems with singularities or degeneracies, the outcomes of such encirclements can be dramatic. The degeneracies of non-Hermitian systems are the exceptional points, which arise when the eigenvalue and eigenvectors of the Hamiltonian coalesce simultaneously. Depending on the number of eigenvalues and eigenvectors coalescing, we decide the order of the exceptional point (typical behavior of complex energy eigenvalue spectrum of a two-band non-Hermitian Hamiltonian is depicted in Figure 5 with the detuning range chosen to exhibit the occurrence of the exceptional point). When a second-order exceptional point is encircled once, the eigenstates get swapped while only one of the eigenstates accumulates the Berry phase of pi radians. This behavior is a resultant of the branch-point character of the complex eigenfrequencies and the associated topologies. Such peculiar phenomena lead to a fractional topological charge, which is peculiar to non-Hermitian systems.

\subsection{Nonlinear Topological Photonics}

The Association of topological behavior and nonlinear photonics has the potential to enrich our understanding of light propagation. These studies can be taken up in two fundamental ways: in the first approach, we focus on a system with preexisting topological features, and optical nonlinearities are employed to make a phase transition. In contrast, in the second approach, we aim to derive the topological features in a trivial system solely from nonlinear optical effects. The first approach is relatively straightforward, and a few preliminary investigations based on this premise have been reported. These studies grapple with the question of whether the incipient optical nonlinearities in a system can cause a change in the system’s topological characteristics. The idea is to design systems where below a certain power threshold, the system remains in some topologically trivial state while above it starts displaying the characteristic robustness associated with the topological propagation~\cite{maczewsky2020nonlinearity,smirnova2020nonlinear}
We envisage that a combination of topological photonic structures and nonlinear effects can provide access to elusive features like nonreciprocity, topological Mott insulators, and non-Abelian topological insulators. Specifically, nonlinear topological photonics looks promising in generating coveted functionalities like active tunability, self-interaction effects resulting in edge solitons in photonic lattices, and many-body quantum topological phases of light. Apart from this, a considerable potential also exists in designing parametric amplifiers immune to feedback and ultrafast optical switches.

\subsection{Higher Order Photonic Topological States}

The topological insulators that we have been referring to until now are insulating in bulk while possessing gapless boundary modes whose dimension is given by (n-1), ‘n’ being the dimensionality of the topological matter in question. The famous bulk-boundary correspondence relates the number of these modes with the bulk properties. However, recently a new kind of topological phase, called the higher-order topological phase~\cite{kim2020recent}, has been noticed where such correspondence is not respected. The reason being, in such occurrences, topologically nontrivial modes with dimensionality more than one dimension lower than the bulk can exist. For example, a higher-order topological insulator with bulk dimensionality ‘n’ gives rise to topological gapless boundary modes of dimensions (n-2) or even lower, in addition to the more prevalent (n-1) dimension gapless boundary modes. In fact, a jth order topological insulator exhibits gapless boundary states of dimensions (n-j). These possibilities have brought the focus onto the realization of in-gap 0-dimensional corner states in 2D and 3D systems, which can be crucial for providing access to robust cavity modes in scalable photonic systems~\cite{zhang2020higher}.

\subsection {Topological Anderson insulators}

A fundamental characteristic of two-dimensional topological insulators is the display of quantized conductance, which remains immune to the disorder, if any,  in the underlying lattice. However, for a strong enough disorder strength, the bandgap closes, and the topological protection is lost. In this scenario, all the boundary states become localized, and the system becomes a trivial insulator, as mandated by Anderson localization. In this context, a surprising discovery was made when the calculations in the HgTe quantum wells suggested a reverse transition where the exponentially small conductance makes way for a quantized perfect conductance by incorporating the strong disorder~\cite{meier2018observation}. The materials which exhibit this new state have been termed as the Topological Anderson Insulator (TAI). Indeed in many realizations, quantized conductance has been induced instead of being inhibited by incorporating a strong enough disorder in the topologically non-trivial material. This effect has been very recently demonstrated in a photonic waveguide lattice, much due to the relative ease with which artificial engineered lattices can be created in such a system, as mentioned before.

\section{Conclusion}

The last decade has vindicated the fact that photonics can be a key enabler in detailed and systematic investigations of topological aspects and has the potential to spearhead future research prospects in the domain. This becomes particularly important in the studies of engineered systems because the length scales of interest in photonics are at least an order larger than those existing in quantum condensed matter systems. Therefore, we believe that the confluence of photonics and topological physics is poised to see many significant developments and research activity in the years to come and will further enrich our understanding of quantum matter.

\section{Author Information}

Nitish Kumar Gupta (email-nitishkg@iitk.ac.in) is a PhD student at Centre for Lasers \& Photonics, Indian Institute of Technology Kanpur. 
Arun M. Jayannavar (email-jayan@iopb.res.in) is a J.C. Bose National Fellow and a distinguished senior scientist currently associated with Bhaurao Kakatkar College, Belgaum.

\section{Acknowledgement}

Arun M. Jayannavar acknowledges DST, India for J C Bose fellowship.

\bibliography{references_aps}% Produces the bibliography via BibTeX.

\end{document}